# Magnetic and Optical properties of strained films of multiferroic $GdMnO_3$

Mohammed S. Al Qahtani, Marzook Alshammari, H.J Blythe, A.M. Fox and G.A. Gehring

Department of Physics and Astronomy, University of Sheffield, Sheffield S3 7RH, UK

N. Andreev, V. Chichkov, and Ya, Mukovskii

National University of Science and Technology "MISiS" , Leninskii Prospect 4, 119049 Moscow. Russia

## Abstract

The effects of strain on a film of mulitferroic $GdMnO_3$ are investigated using both magnetometry and magneto-optic spectroscopy. Optical spectra, in the energy range 1.5eV – 3.5eV, were taken in Faraday geometry in an applied magnetic field and also at remanence. This yielded rich information on the effects of strain on the spin ordering in these films. Epitaxial films of $GdMnO_3$ were grown on $SrTiO_3$ and $LaAlO_3$ substrates. The $LaAlO_3$ was twinned and so produced a highly strained film whereas the strain was less for the film grown on $SrTiO_3$. The Neél temperatures and coercive fields were measured using zero field data and hysteresis loops obtained using a SQUID magnetometer. Optical absorption data agreed with earlier work on bulk materials. The two well known features in the optical spectrum, the charge transfer transition between Mn *d* states at ~2eV and the band edge transition from the oxygen *p* band to the *d* states at ~3eV are observed in the magnetic circular dichroism; however they behaved very differently both as a function of magnetic field and temperature. This is interpreted in terms of the magnetic ordering of the Mn spins.

## I Introduction

In recent years there has been great interest in the physics of "multiferroics". Discovered in the 70's, these unusual materials exhibit ferroelectric and magnetic orderings simultaneously. These two order parameters are strongly coupled by a magnetoelectric coupling, allowing the reversal of the ferroelectric polarization by the application of a magnetic field or the control of the magnetic order parameter by an electric field. Together with recent unprecedented progress in epitaxial thin film growth, this unique property makes these "old" materials very good candidates with which to build multifunctional spintronic devices [1,2]. The multiferroic transition in the orthorhombic manganites, $RMnO_3$ where R= Pr, Nd, Sm, Eu, Gd, and Tb is currently understood with the help of two key ingredients. First, magnetic frustration due to competing exchange integrals between successive neighbours stabilizes a spiral magnetic phase below the Néel temperature $T_N$ [3] although recent work suggests that there may be C-type antiferromagnetic spin order in which the Mn spins are antiferromagnetically arranged in the *a-b* plane but stacked ferromagnetically up the c –axis [4]. Next, in order to lower the spin-lattice coupling energy the oxygen atoms are pushed off the Mn-Mn bond, driving an electric polarization at $T_N$. Charge, spin, orbital, and lattice subsystems are linked very tightly and so we investigate the effects of strain in thin films caused by lattice mismatch to different substrates on the magnetic properties of one particular manganite, $GdMnO_3$.

The magnetic phase diagram has been mapped out for bulk crystals [5,6] as a function of temperature and magnetic field and also as a function of pressure. It is found that as the temperature is lowered the first transition occurs at 40K and is a transition between the paramagnetic phase and an antiferromagnetic phase. In zero magnetic fields there is a transition to a canted phase at ~18K and that the Gd spins order at close to 7K. However below ~20K there is strong hysteresis so that several phases can coexist [6]. The phase boundaries depend both on the magnitude of the magnetic field and its direction. In particular a ferroelectric phase appears below ~7K for a magnetic field greater than ~2T along the *b* axis. The phase diagram is very dependent on pressure [7,8] as expected due to the existence of a ferroelectric phase. The first phase transition was found to be shifted to 44K for nanoparticles with mean size of 54nm [9].

The optical absorption is particular interesting for these manganites [10, 11] because of the suppression of the clear peak seen in the optical absorption at ~2eV seen in $LaMnO_3$ due to a charge transfer excitation between Mn ions which has been related to the different canting of the Mn *d*-orbitals and also on the directions of Mn spins. We use magneto-optic spectroscopy because it weights the strength of the observed transitions by the amount by which magnetic states are involved. Magneto-optical studies on single crystals of $TbMnO_3$ in a magnetic field have shown a strong feature occurring near 3eV [12].

We report on measurements made on thin films of $GdMnO_3$ deposited on two different substrates using RF microwave sputtering. $SrTiO_3$ expands the basal plane of $GdMnO_3$ and hence gives a contracted c axis and a twinned substrate of $LaAlO_3$ which compresses and distorts the film in the basal plane and expands in along the c axis. Magnetic measurements are made in a SQUID magnetometer. The optical absorption is measured at room temperature and the Faraday rotation and magnetic circular dichroism (MCD) are measured as a function of temperature both in field and at remanence. Finally the results are discussed.

## II Sample Preparation and Characterisation

Targets of $GdMnO_3$ were prepared from stoichiometric mixtures of $Gd_2O_3$ and $MnO_2$. Films of thickness of 100-200nm of $GdMnO_3$ were sputtered using a mixture of Ar and $O_2$ at a pressure of 1-2 mTorr on to a substrate which was heated to $650^0$C. This produced an epitaxial film with the [110]-direction parallel to the plane of the substrate. X-ray data taken at room temperature is shown in Fig. 1. The $LaAlO_3$ substrate was found to be heavily twinned however the $GdMnO_3$ film was found to have grown epitaxially in both cases. The $SrTiO_3$ substrate undergoes a structural transition at ~100K; this may produce further strains on the film in the temperature range where the magnetic ordering occurs.

## III Magnetic Measurements

The magnetic properties of the films were investigated by SQUID magnetometry with the magnetic field in the plane of the samples. Field cooled and zero field cooled magnetisation plots are shown in Fig.2. The data was taken in a field of 100Oe and the contribution from the diamagnetic substrates has been subtracted. The transition to the antiferromagnetic phase at 47K in bulk shows up clearly in these plots however it is shifted to lower temperature. It occurs at 42K for the films on the $SrTiO_3$ substrate and at 39K for the film on the $LaAlO_3$ substrate. These should be compared with the temperature of 44K observed for nanoparticles [9]. This indicates that the strain induced by the substrates is more significant than the finite size effects in reducing $T_N$; the exchange interactions between the Mn ions is very sensitive to strain [3]. The lower transition to a canted phase, observed at ~18K in bulk, is not visible in this plot.

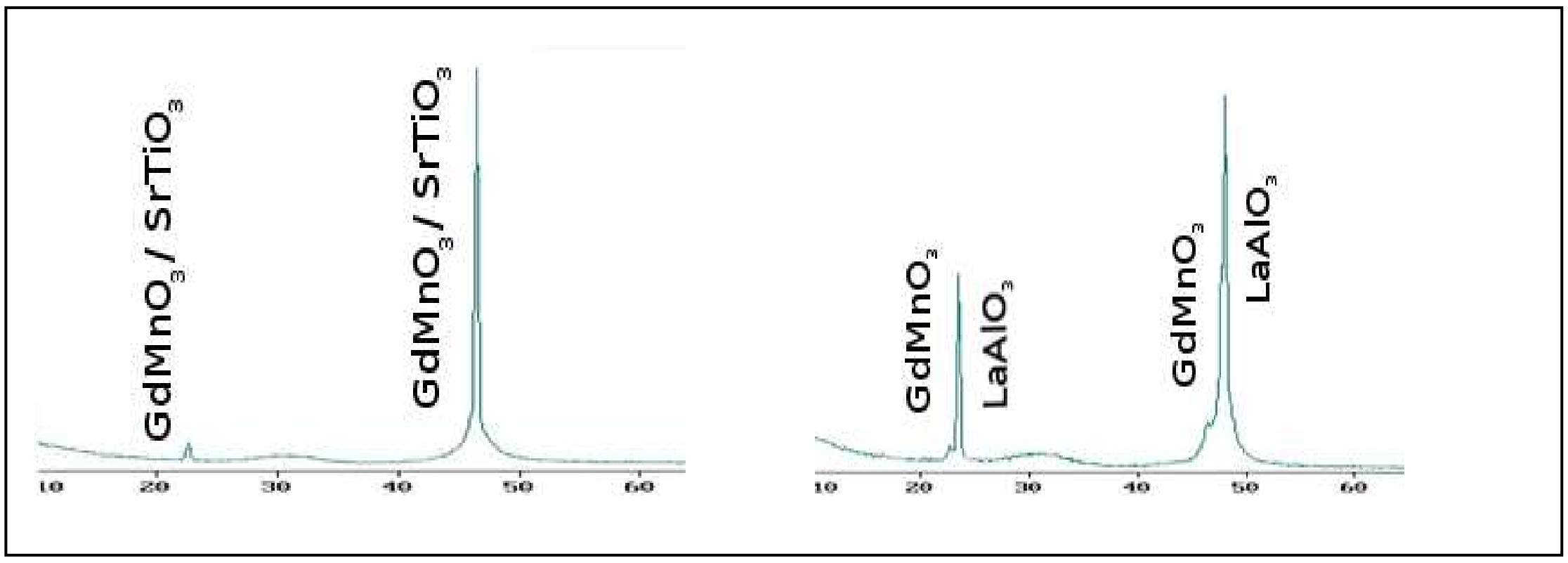

Fig.1 Xray data from films of $GdMnO_3$ on the $SrTiO_3$ and $LaAlO_3$ substrates showing epitaxy.

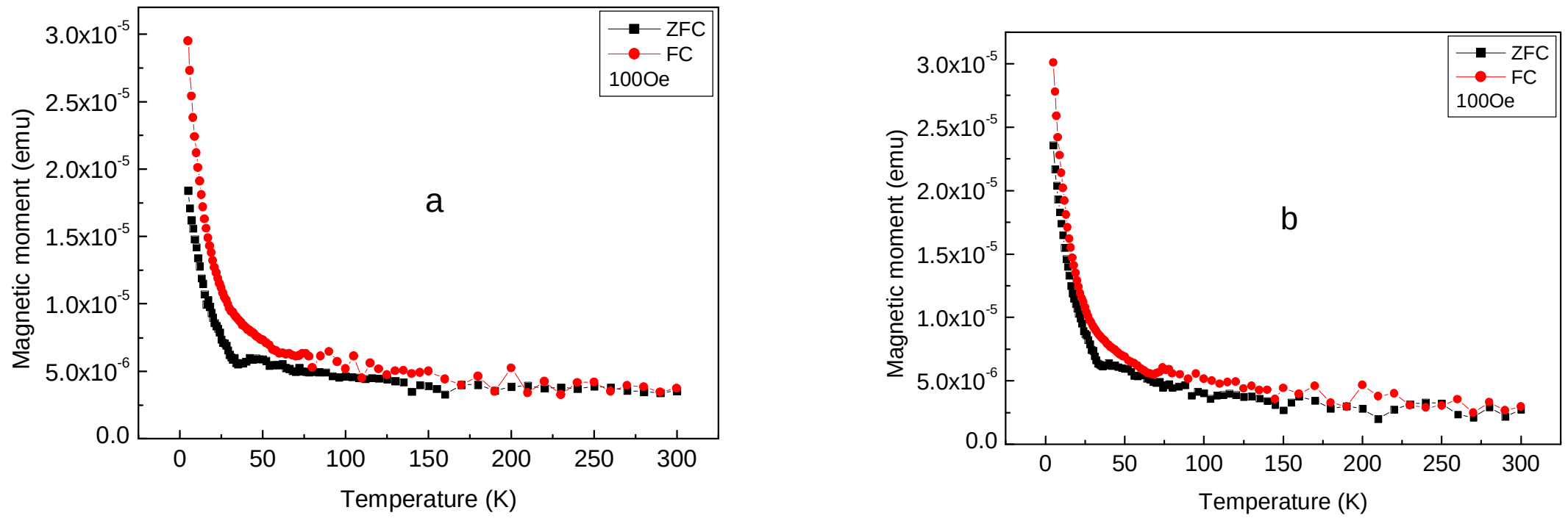

Fig. 2 (Colour on line) Field cooled and zero field cooled data taken in an in-plane field of 100Oe for films of $GdMnO_3$ on (a) $SrTiO_3$ and (b) $LaAlO_3$ substrates. The data for the bare substrates has been subtracted.

A coercive field should appear as we enter the canted phase however we have only detected a sizeable coercive field at 5K.

We took hysteresis loops with the field in plane and subtracted off a contribution that varied linearly with the field in the range 7kOe -10kOe. The subtracted contribution was due to the diamagnetic signal from the substrate and a paramagnetic signal from the film – this was large at low temperatures following the low field susceptibility shown in Fig. 2 and is shown in Fig. 4. In Fig. 3 we show the loops at temperatures 5K and 10K, the moment is higher at 5K in the film on $LaAlO_3$; there was no clear signal at 20K or higher.

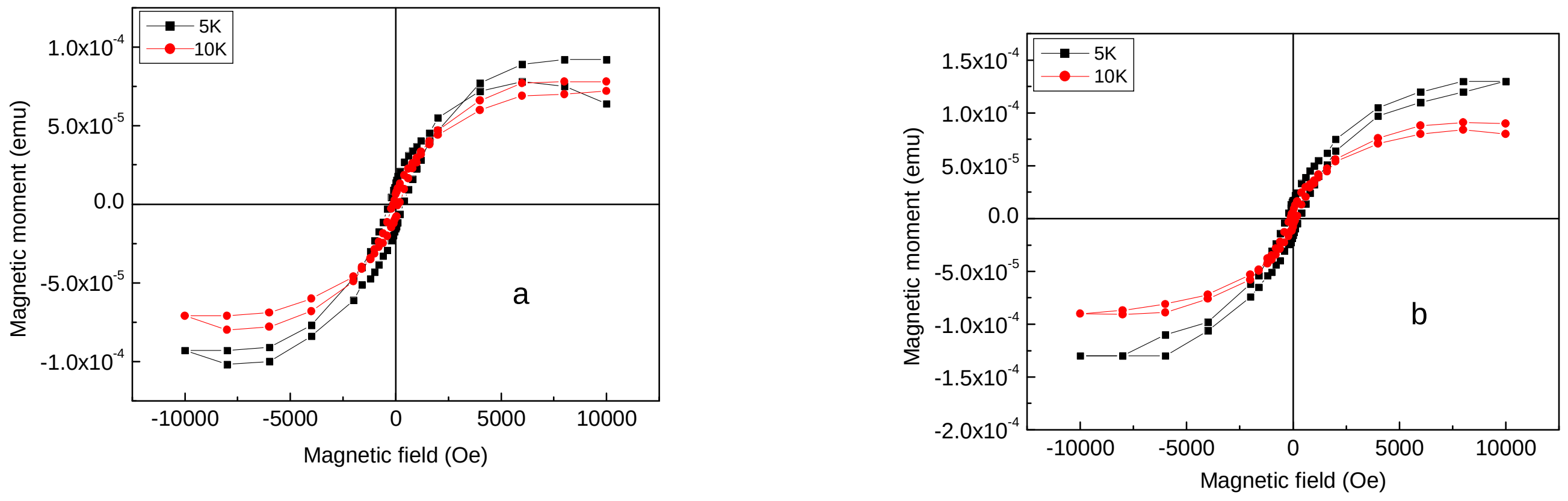

Fig. 3 (Colour on line) The hysteresis loops for the field in plane (with linear background subtracted) at temperatures 5K and 10K for the films on (a) $SrTiO_3$ and (b) $LaAlO_3$ substrates.

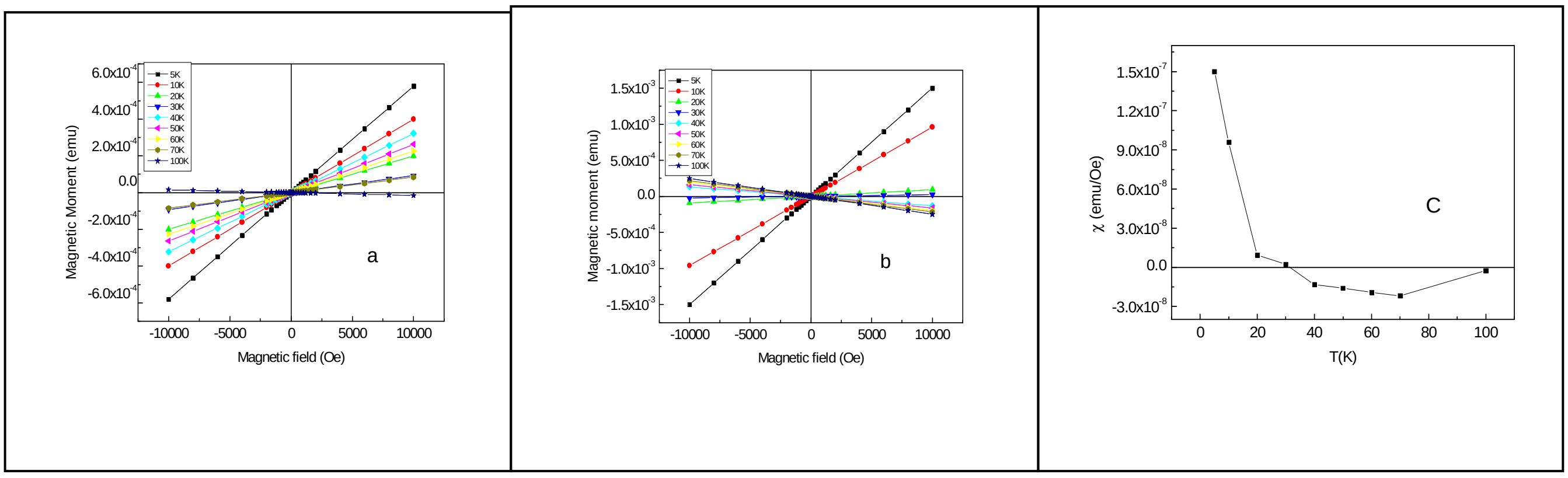

Fig. 4 (Colour on line) The subtracted susceptibility of the films (a) on the $SrTiO_3$substrate (b) on the $LaAlO_3$substrate and (c) the high field susceptibility for the film on the $LaAlO_3$substrate.

## IV Optical Measurements

The absorption spectra of the two films were measured at room temperature. The absorption for the sample on $LaAlO_3$ behaves as expected; it rises smoothly from a gap at ~ 3.48 eV up to the highest energy measured, 4.5eV. This is due to charge transfer transitions from the oxygen 2*p* band to the unoccupied Mn *d* states. These states are in the higher $e_g$ orbital but with spin parallel to that of the electron in the occupied $e_g$ orbital [11]. The data for the film on the $SrTiO_3$ substrate is only meaningful below 3.2eV because the absorption at higher energies is dominated by the substrate absorption edge [13]. The absorption at ~2eV seen in the optical absorption in $LaMnO_3$ which is due to a charge transfer excitation between Mn ions is strongly suppressed in these measurements due to the canting of the Mn *d*-orbitals [10,11,12].

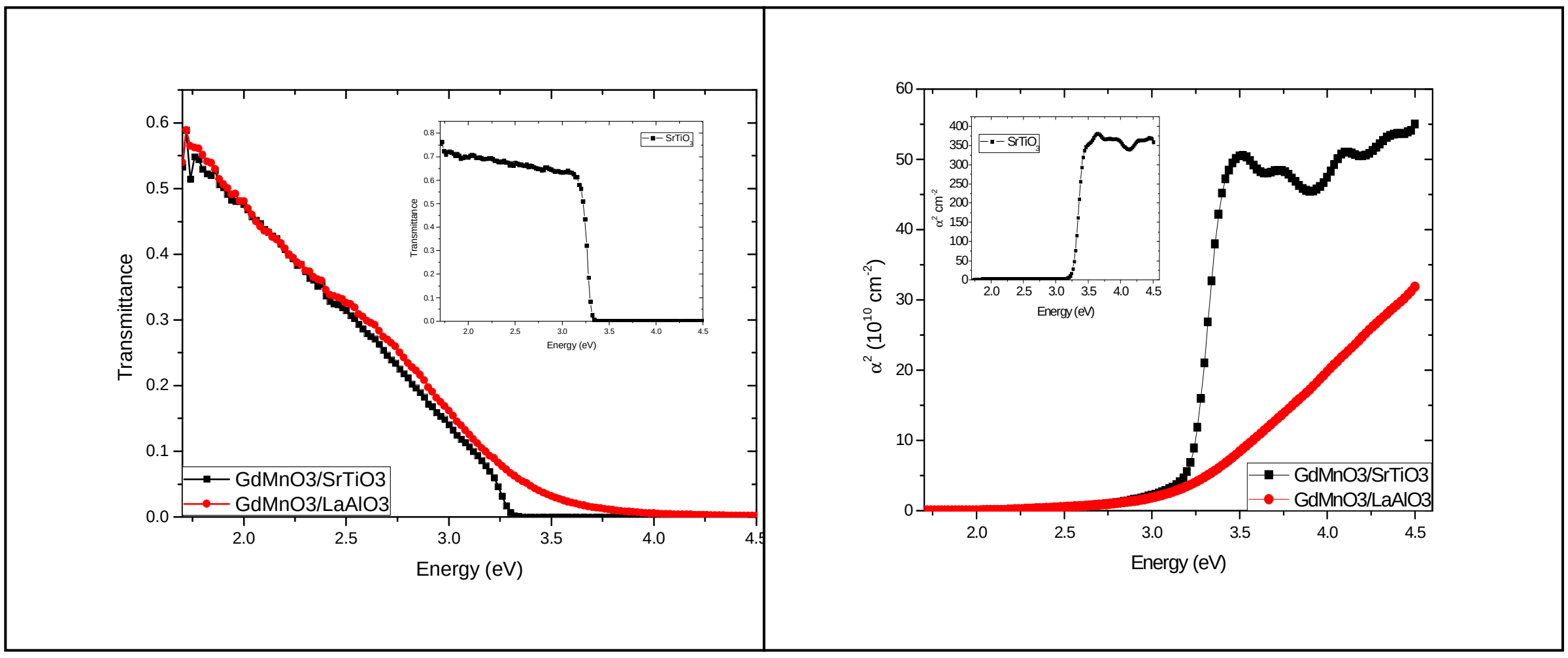

(a) (b)

Fig. 5(Colour on line) (a) Transmittance (b) the absorption $\alpha$ squared for the two films. The $SrTiO_3$ substrate absorbs strongly above 3.2eV and its spectra are shown in the inserts.

oxygen 2*p* band to the Mn *d* states. The data for the film on the $SrTiO_3$ substrate is only meaningful below ~3.2eV because the spectra at higher energies are dominated by the substrate absorption edge [13]. The absorption at ~2eV seen in the optical absorption in $LaMnO_3$ which is due to a charge transfer excitation between Mn ions is strongly suppressed in these measurements due to the canting of the Mn *d*-orbitals [10,11,12].

As shown in [9,10] the matrix element for the transition between two manganites ions depends on the symmetry of the ground state orbital of one ion and the excited state of a neighbouring ion and their relative spin orientation. The orbital angle $\theta$ is defined in terms of the relative strength of the local tetragonal and orthorhombic distortions $q_1$ and $q_2$ respectively as $\tan\theta = q_2/ q_1$ has been measured recently for various rare earths $\theta \sim 117^0$. The temperature dependence of the incommensurate wave vector characterising the spin ordering has been measured for $TbMnO_3$ [3], it is temperature dependent and varies between 4/7 and 11/20 in units of the inverse lattice vector $b^{-1}$. This means that the neighbouring spins are close to being antiparallel and this will also further inhibit the charge transfer transition at ~2eV at low temperatures.

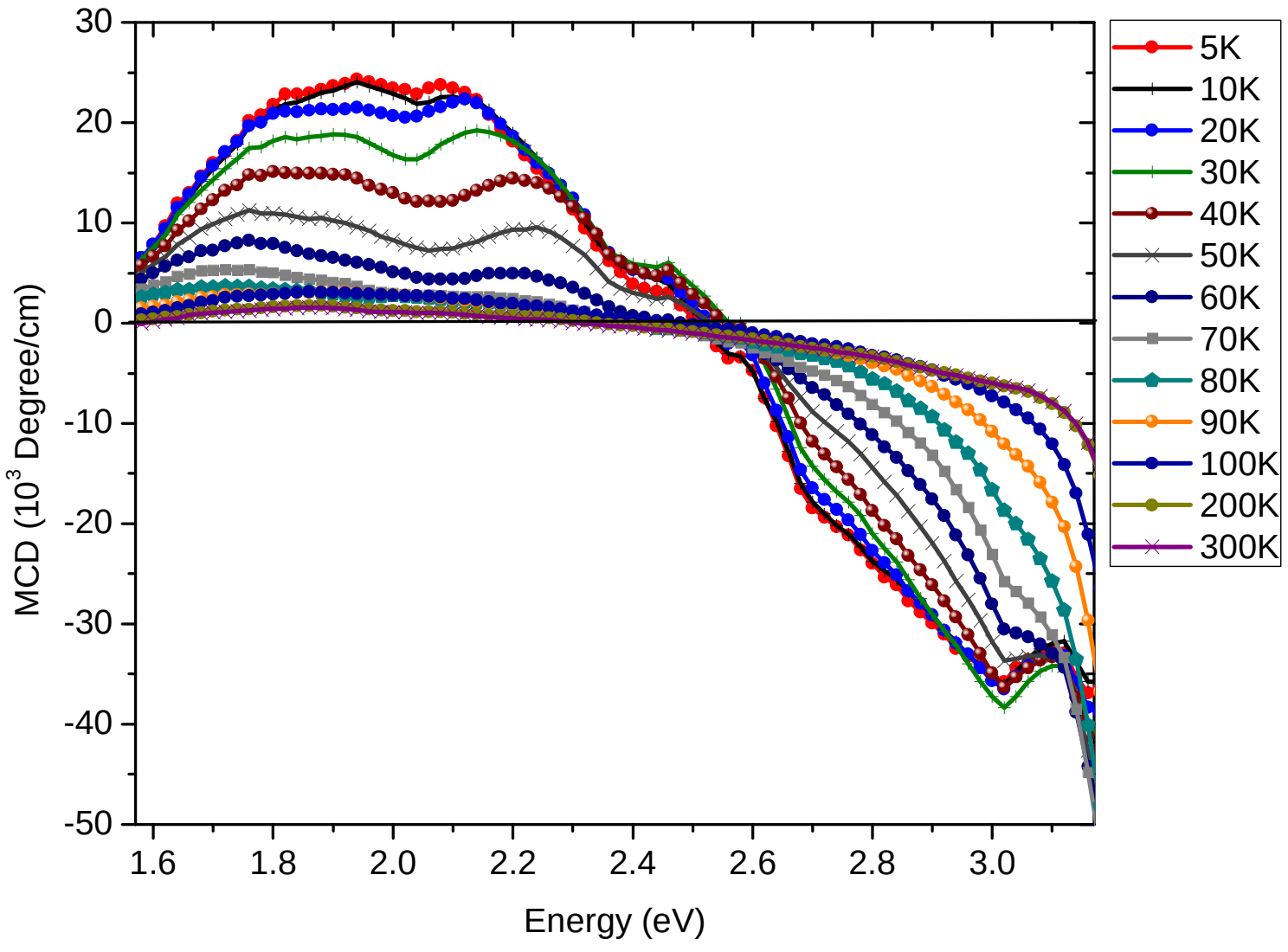

Fig. 6 (Colour on line) The temperature dependence of the MCD of $GdMnO_3$ taken at 0.5T as a function of temperature. The contribution from a blank $SrTiO_3$ substrate has been subtracted from this data.

We now consider the magneto-optical spectra in Faraday geometry where the magnetic field is applied normal to the plane of the film, along the *c*-axis. The $SrTiO_3$ substrate undergoes a structural transition at ~100K which gives rise to a birefringence unless the $E$ vector of the light is along the optic axis [14]. We were careful to orient our samples so this effect was not a perturbing feature. The difference in absorption between left and right circularly polarised light is the magnetic circular dichroism (MCD) which is directly proportional to the imaginary part of $\varepsilon_{xy}$, it occurs because of an imbalance between transitions in which the orbital angular momentum changes by ±1. There are two ways in which this can occur. The two transitions can have different intensities because of differing occupations of either the initial or the final state; in this case the MCD spectrum mirrors the shape of the absorption spectrum. The other possibility is that the two transitions occur with equal intensities but at slightly different energies and in this case the MCD spectrum has a dispersive feature where the absorption spectrum has a peak. In $GdMnO_3$ the MCD is of the former type because the allowed transitions are dominated by the spin occupations of the Mn *d* orbitals. Transitions to unoccupied *d* levels with opposite spin are higher in energy by the Hubbard energy, $J_H$ ~1eV, and so would appear at different parts of the spectrum.

We consider two sets of data. In the first, shown in Fig. 6, the data was obtained from the difference in the induced ellipticity of the light for data taken in fields ±$B$ where $B$=0.5T: this is a standard MCD spectrum. In the second set of data we measured the spectrum in zero magnetic field after having reduced the field to zero from values of ±0.5T, this is the remnant spectrum and is shown in Fig. 7.

These results taken together are really surprising because the spectral shapes of the MCD spectra taken in field and at remanence are so dissimilar, which is, as far as we know, a unique occurrence. One expects that the remnant spectrum has the same shape as the spectrum taken in field but with a magnitude scaled by the ratio of the remnant magnetisation to the applied magnetisation. The peak at ~2eV at 10K in the remnant spectrum has been reduced by a factor of ~5 relative to its value in field; this is comparable to the ratio of $M_r/M_s$ observable in the SQUID data shown in Fig. 3.

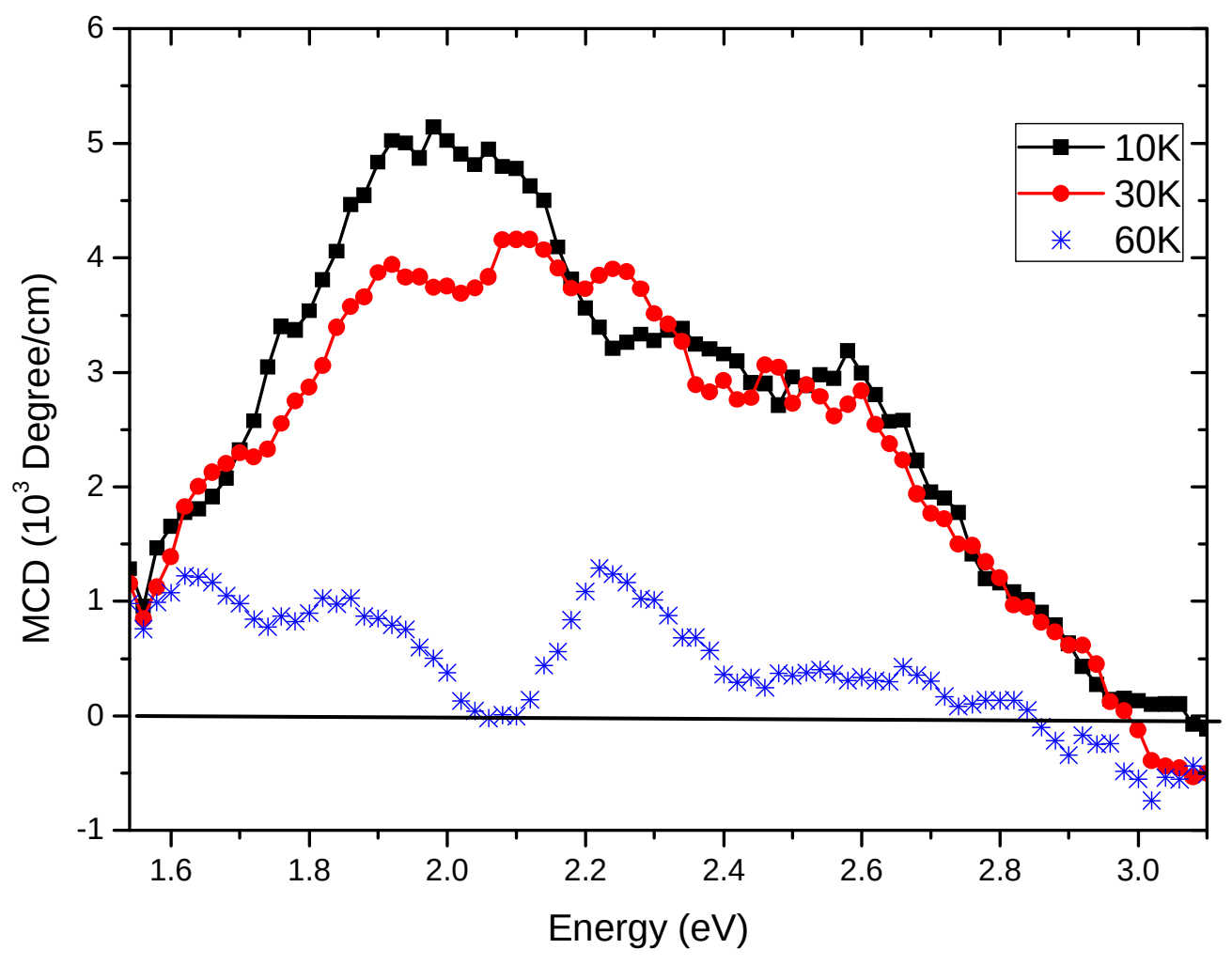

Fig. 7 The MCD spectrum taken at remanence for $GdMnO_3$ on $SrTiO_3$ there is a finite spectrum at 10K and 30K, the spectrum at 60K is almost zero. The remanent spectra for a blank substrate (which was anyway very small, < 0.1) has been subtracted from this data.

This data can be understood when it is realised that there are two distinct components of the spectra with different temperature and field dependencies, these are the negative feature occurring above 3eV and a positive peak occurring near 2eV.

The negative feature occurs near the charge transfer band edge and it changes rather little from 5k to 40K in spite of the fact that the susceptibility is rising strongly over this temperature range, and it is visible out to high temperatures. This suggests that the signal depends on the existence of a finite magnetic susceptibility for a field along *c* but is rather independent of the canting that occurs below 18K. This shape is reminiscent of the absorption and occurs if the strength of the absorption is stronger for one spin direction than the other. This is expected to occur for the transition from the oxygen-*p* band to an unoccupied Mn *d* state when the Mn states are polarised by a magnetic field because the spin of the lowest unoccupied state will be parallel to the occupied state [10]. A similar feature was seen in the imaginary part of $\varepsilon_{xy}$ in $TbMnO_3$ although in this case the field was applied along *b* [12], and they could measure to 4eV and see a peak whereas our data was cut off by the absorption of the $SrTiO_3$ (they also used an opposite sign convention). This effect vanishes at when the magnetic field is zero.

The energy of the transition at about 2eV is characteristic of the charge transfer transition between neighboring Mn ions and depends on the overlap of the spin states. For spins oriented at an angle $\varphi$ to each other this varies as $\cos^2\varphi/2$. The strength of this peak in the MCD spectrum, shown in Fig. 6, decreases much more rapidly with temperature, than the strength of the peak at 3eV. For example there is a significant drop in the intensity of the 2eV peak at 30K relative to 5K and it has been suppressed almost to zero at 60K whereas the intensity of the 3eV peak at 30K is changed very little compared with the data taken at 5K and the intensity at 60K has only dropped by a factor of ~2. This is consistent with the strength of the 2eV peak depending on the relative spin directions of two neighbouring Mn ions and hence the square of the magnetisation. On the other hand it is this peak that survives in the remanence spectrum. One would expect a finite remnant spectrum only when there is magnetic hysteresis which should occur below the transition to the canted phase at 18K and so it is very surprising to obtain such a high remanence spectrum at 30K. It is possible that the strain in the film is inducing a canted phase already at $T_N$. The peak in $LaMnO_3$ that was suppressed at ~2eV is in fact 3 peaks at 1.9eV, 2.3 eV and 2.6 eV but this was not apparent in the absorption data. The MCD intensity is related to the absorption but weighted by the amount by which the transition involves the magnetic Mn ions which suggests that these transitions should be more visible in the MCD than in the absorption. There are features in the remanence spectra that might correspond to these peaks but this is not certain. It is clear that the 2eV peak can be studied much better using MCD at remanence than in any other mode because other features are absent.

## IV Conclusion

We have found considerable differences between the magnetic behaviour of bulk $GdMnO_3$ and that grown on the two substrates, $SrTiO_3$ and the twinned substrate of $LaAlO_3$. There is a structural transition in $SrTiO_3$ at ~100K and this may be straining the film grown on this substrate at low temperatures. The Neél temperature is reduced and there is some evidence that a canted phase appears with the onset of antiferromagnetism.

This report given has shown the magneto-optic data on thin films of $GdMnO_3$ grown on $SrTiO_3$ taken in a field of 0.5T and at remanence. A dramatic result was the stark difference between the shapes of these two spectra. This was understood in terms of the two components giving rise the optical spectra, the charge transfer transition between Mn ions occurring near 2eV and the charge transfer between the valance band of oxygen *p* states and the unoccupied Mn *d* orbitals. The MCD spectra indicated that the transition involving the *p* states scaled with the magnetisation and could be ascribed to the spin dependence of the allowed transition arising from the occupation of the Mn *d* orbitals. The transition near 2eV was more strongly dependent on temperature but much less sensitive to the external field and was observed in the remnant spectrum. It was a clear feature in the MCD whereas it is strongly suppressed in the optical absorption this is because it depends so strongly on the spin dependence of the occupation of the Mn *d* orbitals. It depends on the relative orientation of the spins of the neighbouring Mn ions. It is clear that study of the MCD spectrum of this material is a rich source of information that is not available from a study of the absorption alone.

**Acknowledgments**

We should like to acknowledge that the work in Sheffield was supported for a studentship for M Alshammari from K.A.C.S.T. (Saudi Arabia) and the work in Moscow was supported by Joint RFBR-NSF Grant 09-02-92661 and NA was supported by the Programme of Creation and Development of the NUST "MISiS".